\documentstyle[psfig,floats,twocolumn,aps,prl]{revtex}
\input rotate
\newbox\rotbox
\begin{document}
\title{\vspace*{-1cm}\hfill
 {\tt Submitted to Phys.~Rev.~Lett.}
       \vspace{1cm}\\
Scattering of rare-gas atoms at a metal surface:
evidence of anticorrugation of the helium-atom
potential-energy surface and the surface electron density}
\author{M.~Petersen, S.~Wilke,\cite{aaa} P.~Ruggerone, B.~Kohler, and
M.~Scheffler}
\address{
Fritz-Haber-Institut der Max-Planck-Gesellschaft \\ Faradayweg 4-6, D-14\,195
Berlin-Dahlem, Germany.}
\date{submitted October 10, 1995}
\twocolumn[
\maketitle
\begin{quote}
\parbox{16cm}{\small
Recent measurements of the scattering of He and Ne atoms at Rh\,(110)
suggest that these two rare-gas atoms measure a {\em qualitatively} different
surface corrugation: While Ne atom scattering seemingly reflects
the electron-density undulation of the substrate surface, the scattering
potential of He atoms appears to be anticorrugated. An understanding of this
perplexing result is lacking. In  this paper we present density functional
theory calculations of the interaction potentials of He and Ne with
Rh\,(110).
We find that, and explain why, the nature of the interaction of the two
probe particles is qualitatively different, which implies that
the topographies of their scattering potentials are indeed anticorrugated.
}
\end{quote}]
PACS numbers:79.20.Rf, 68.35.Bs, 73.20.-r\\

The presumed simplicity of the interaction of low energy rare-gas atoms with
surfaces has strongly promoted the use of He atoms as
{\it ideal probe particles} in scattering experiments in order
to determine the surface atomic geometry and the corrugation
of the surface electron-density~\cite{Eng82,Hup92,Rie94}. Reconstructed
surfaces
have been successfully investigated
and even light
adsorbates like hydrogen, which are scarcely seen with other techniques,
can be located.
As a consequence, helium atom scattering (HAS)
is one of the leading tools for the analysis of structural and vibrational
properties of surfaces.
A widely used form for the interaction potential has been derived by Esbjerg
and N{\o}rskov~\cite{EsNo80} who argue
that the interaction energy of a He atom and a surface
is simply proportional to the unperturbed electron density of the substrate
at the position of the He atom.
Although widely accepted,
the credibility
of this approach has been questioned: For Ni\,(110) Rieder and Garcia
\cite{RieGa82} reported a
serious discrepancy in the corrugation amplitude
when comparing their measurements with the results
of the Esbjerg-N{\o}rskov approach taken together with good quality
calculations of the surface electron density.
Annett and Haydock \cite{AnHay84a}
tried to reconcile the disagreement by introducing an additional
term in the interaction potential. This addition, named
{\it anticorrugating term}, arises from the
hybridization between the occupied $1s$ orbital of the He atoms and the
unoccupied states of the metal surface and results in an attractive
contribution to the potential which should be stronger at on--top positions
than at bridge sites.
Harris and Zaremba~\cite{Har85a} criticized the estimates by Annett and
Haydock
and argued that the {\em anticorrugating term} should be by more than an
order
of magnitude smaller than what Annett and Haydock had evaluated.
Harris and Zaremba~\cite{Har85a} claimed that the discrepancy
between theory and experiment lies in an improper description of the
van der Waals contribution and a tendency to over-corrugate the He-surface
interaction potential within the framework of the local-density approximation
(LDA) of the exchange-correlation interaction. A general consensus on the
origin of the effect noted by Rieder and Garcia \cite{RieGa82} has not
been reached.

Recent measurements~\cite{Rie93}  enforced the interest in
this important question and in fact raised significant doubts about the
meaning and interpretation of the important HAS method. Rieder and coworkers
found unexpected differences in the interaction potentials and measured
corrugations when comparing the scattering of He and Ne atoms at surfaces.
For Rh\,(110) and Ni\,(110) and using the Esbjerg-N{\o}rskov
approach they concluded that the Ne diffraction data reflect the corrugation
of the surface atomic structure and the unperturbed electron density.
In the case of He atom scattering, however,  the same type of analysis
gave an electron corrugation shifted away from the atomic positions:
The electron density at the short--bridge position appeared in HAS
to be higher above the surface than the on--top site.
Rieder's explanation, following the arguments of Annett and Haydock
\cite{AnHay84a}, is that especially at the on--top position the He $1s$
orbitals, as well as the Ne $2s$ orbitals, and the empty metal $s$ states
hybridize giving rise to an anticorrugating contribution. In the case of Ne,
however, this contribution
is overcompensated by the repulsive interaction between the Ne $2p_{x,y}$
orbitals and the metal $s$ states. Severe doubts about this explanation
are in place because it assumes that the additional term introduced by Annett
and Haydock is now even dominating the interaction potential.

Obviously, there is  profound need for a direct calculation of the interaction
of a rare-gas atom with a metal
surface. For this it is important that practically no serious constraints
on the electronic response of the surface and  the inertness of
the rare-gas atoms are introduced.
A related important aspect of such theoretical study is
that the interaction of a He or Ne atom
with a metal surface is an example of weak physisorption, and a calculation
represents a critical test of the  exchange-correlation functional
used in {\em ab initio} calculations.

We performed systematic density-functional-theory (DFT) calculations
exploiting two different functionals for the exchange-correlation
interaction,
namely the LDA~\cite{cep} and the generalized
gradient approximation (GGA)~\cite{GGA}.
If not stated explicitly, the below reported results refer to a DFT-GGA
calculation. The Kohn-Sham equations and energy
functionals are evaluated self-consistently using the full-potential linear
augmented plane
wave (FP-LAPW) method~\cite{Bla93,Koh94}.
The Rh\,(110) surface is treated in terms of a super-cell approach,
using  five layers thick slabs, which are separated by a vacuum region of
18~\AA. The slab thickness is rather small for a (110) surface, but
because of the weakness and localization of the interaction it
is sufficiently large for the present study.
The energy cutoff for the LAPW wave functions is chosen to be
$E_{\rm{cut}}$~=~15.5~Ry, the muffin tin radius $R_{\rm{MT}}$ is 1.24~\AA, the
angular momenta of
wave functions inside of the muffin tin spheres are taken up to
$l_{\rm{max}}$~=~10.  The muffin tin radius for the He and Ne atoms is
$R_{\rm{MT}}$~=~0.9~\AA. For the potential expansion we use a plane-wave
cutoff
of 70~Ry
and a ($l,m$) representation (inside the muffin tin spheres) with
$l_{\rm{max}}$~=~4. The {\bf k}-integration is performed on an equally spaced
mesh of 88 points in the whole two dimensional surface Brillouin zone of a
$(1 \times 1)$ surface cell. For the evaluation of the potential
energies of impinging He and Ne atoms we use a $(1 \times 2)$ surface cell.
It is interesting to note that, because of the small size of the probe atoms
compared to Rh and the weakness of the perturbation,
a calculation with a $(1 \times 1)$ cell gives practically the same results.
All calculations are performed  non-relativistically.

As a first test of the accuracy of the calculations we studied the equilibrium
structure of Rh bulk and the clean Rh\,(110) surface. The theoretical lattice
constant ($a_0^{\rm{th}}$~=~3.89~\AA, without accounting for
zero point vibrations) agrees well with that measured at room temperature
($a_0^{\rm{exp}}$~=~3.80~\AA \cite{Schroe72}).
We find the first layer to relax inwards by $\Delta d_{12}/d_0~= -4.9$~\%,
and the second layer relaxes outwards by $\Delta
d_{23}/d_0~= +2.3$~\%, with $d_0$ being the inter-layer
distance in the bulk. These results are in good agreement
with the values obtained by a LEED analysis ($\Delta d_{12}/d_0~=~-6.8$~\%,
$\Delta d_{23}/d_0~=~+1.9$~\%~\cite{Put89}).

The interaction between the metal surface and the incoming rare-gas atoms is
studied within the adiabatic approximation. The substrate geometry is kept
frozen in the scattering event since the Rh atoms are much heavier.
We also performed calculations with an adiabatically optimized substrate
geometry,
which resulted in negligible changes:
The relaxation of the first layer varied by less than 0.1~\% if a Ne
atom impinges at the on--top position.

\begin{figure}[t]
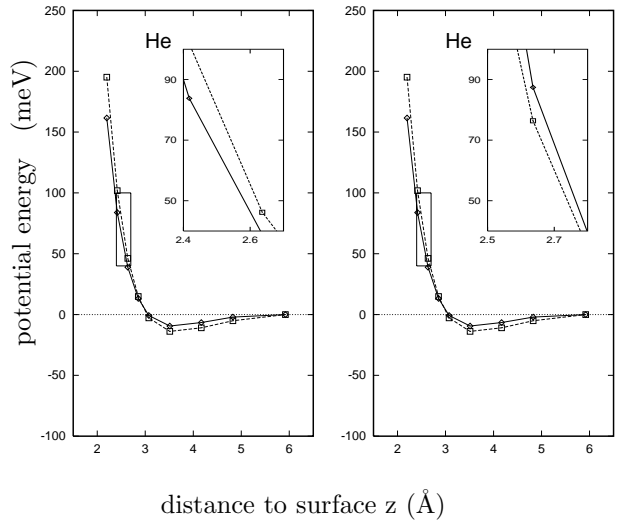

\centerline{\hbox{
\psfig{figure= en.he,height=6cm,angle=-90}
\hspace{-.5cm}
\psfig{figure= en.ne,height=6cm,angle=-90}}}
\begin{center}
\begin{picture}(0,0)
\put(-60,0){distance to surface z (\AA)}
\put(-120,60){\setbox\rotbox=\hbox{ potential energy \ \  (meV)}
\rotl\rotbox}
\put(25,190){ \includegraphics{zoom.ne}}
\put(-90,190){ \includegraphics{zoom.he}}
\end{picture}
\end{center}
\caption{Calculated potential energy using DFT-GGA (see text)
for a He  (left) and Ne (right) approaching the on--top and short--bridge
positions of Rh\,(110) as a function of the distance $z$ from the center of
the first surface layer. The insets show a magnification of the repulsive part
of the potential for particle energies used in experiment. Solid lines are
guides to the eye.}
\label{energy}
\end{figure}

The interaction potential energy was calculated for many positions of the
He and Ne atoms (see Fig.~\ref{energy}). At a distance of 6.0~\AA~ above the
surface we find that the
atom-substrate interaction is negligible and hence use this geometry to
define the energy zero of our calculations.
It was also confirmed that at this distances it makes no difference
where the probe atom is placed parallel to the surface.
The theoretical results for the interaction potential energy
of He (Fig.~\ref{energy}, left) show clearly that the potential energy is
anticorrugated with respect to the
unperturbed substrate electron density and atomic structure: At the repulsive
part and same distance from the surface the energies for the short--bridge
geometries are higher than those of the
on--top geometries. Thus, for the whole range of
particle energies typically used in HAS experiments ($\sim$~20~-~100~meV)
the classical turning point above the on--top position is closer to the
surface
than that above the short--bridge position.
The behavior of Ne nearing the Rh\,(110) surface is qualitatively
different. The potential energy curves  for Ne atoms above the short--bridge
and on--top sites almost coincide, but in
the repulsive regime there is a clear difference and at the same distances
from the surface the short--bridge position is energetically favored over the
on--top position (see Fig.~\ref{energy} right). Thus, according to the
calculations the corrugation experienced by
Ne atom scattering corresponds qualitatively to that of the clean surface
electron
density. This anticorrugation for HAS and ``normal'' corrugation for
Ne atom scattering agrees with the  experimental analysis of
Rieder, Parschau, and Burg~\cite{Rie93}.
Using the data of Fig.~\ref{energy} we estimate the corrugation amplitude
along the [110] direction, $\zeta_{10}^{\rm{scatt}}$, at the
particle energies used by Rieder {\it et al.}~\cite{Rie93}
as the difference in the classical turning point
over the on--top and the short--bridge positions.  We obtain
$\zeta_{10}^{\rm{scatt}} \sim~-$0.06~\AA~ for He and
$\zeta_{10}^{\rm{scatt}} \sim~+$0.04~\AA~  for Ne.
The comparison of these
results with those derived in
Ref.~\onlinecite{Rie93}, $\zeta_{10}^{\rm{scatt}}~=~-$0.04~\AA~ for He and
$\zeta_{10}^{\rm{scatt}}$~=~+0.089~\AA~ for Ne, shows that our results
agree qualitatively and even somehow quantitatively
with those of the experimental analysis.
The quantitative disagreement may be due to the GGA
but it may also
be due to the fact
that the measurements were not performed for a clean Rh\,(110)
but a H-covered surface, since
the adsorbates enabled the identification
of the on--top and short--bridge positions.
Thus, the experimental corrugation amplitude for the clean surface was
extrapolated from the measured data by assuming that the H atoms give rise
to a Gaussian contribution to the clean surface electron density~\cite{Rie93}.

Our DFT-GGA results reproduce not only
the experimental corrugation but are also consistent
with other features of the probe atoms potential energy.
For example, the calculated potential well for a He atom
is 13 meV (8 meV, if we include
the zero-point vibration), which
compares nicely with the value derived from
selective adsorption measurements, 8.2 meV~\cite{Par89}.
For Ne we find 18 meV (11 meV with the zero-point vibration). On the other
hand, it is interesting
that with the LDA exchange-correlation
functional these quantities are in poor accordance with the
experimental data: The turning point for He and Ne are systematically closer
to the surface and the potential wells are too deep (27 meV for He and 61 meV
for Ne). This is consistent with
the well known behavior that the LDA systematically gives rise to an
overbinding in poly-atomic systems.
\begin{figure}[t]
\vspace{-.5cm}
\centerline{\hbox{
\hspace{-1cm}
\psfig{figure=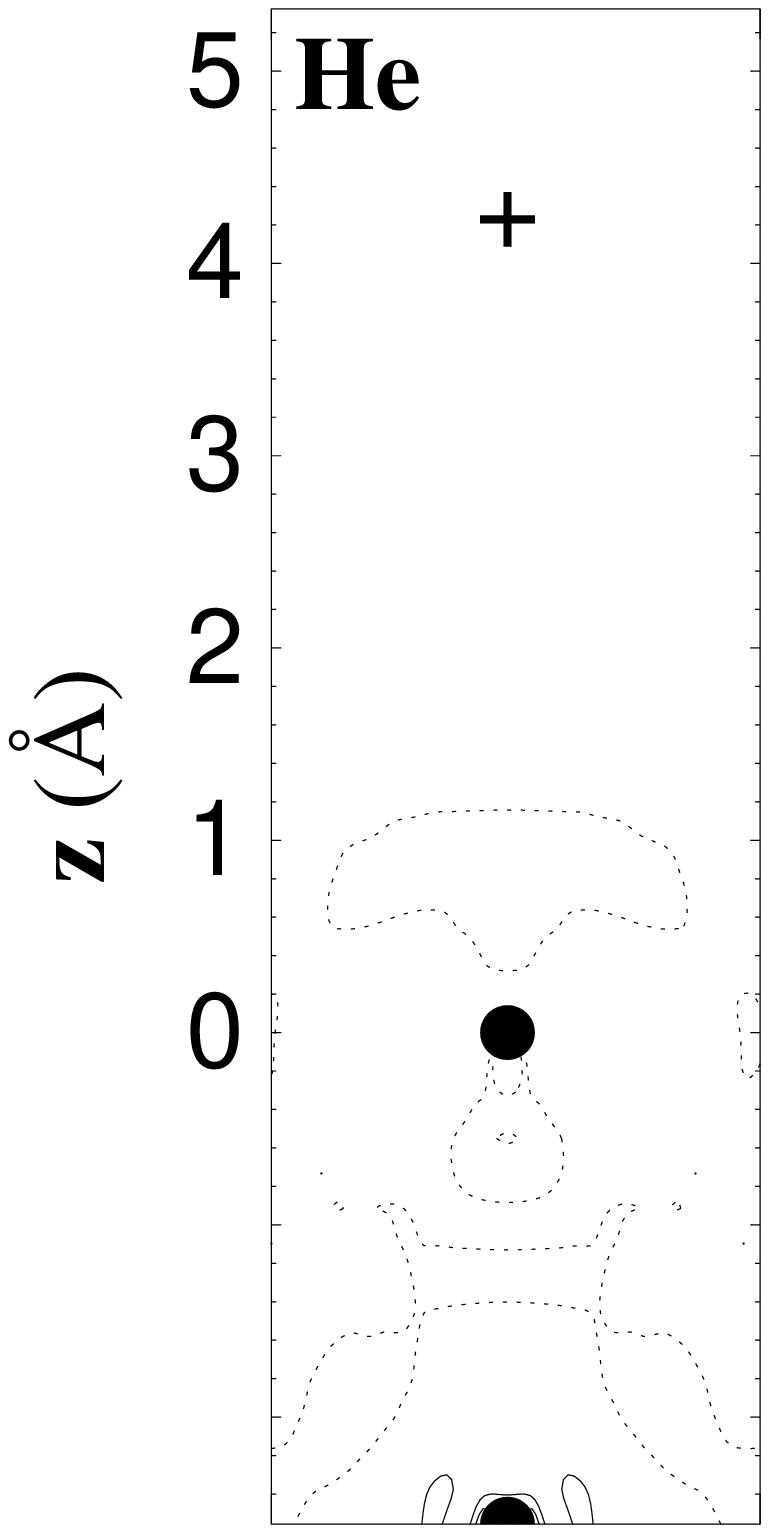,height=7cm}
\hspace{-2cm}
\psfig{figure=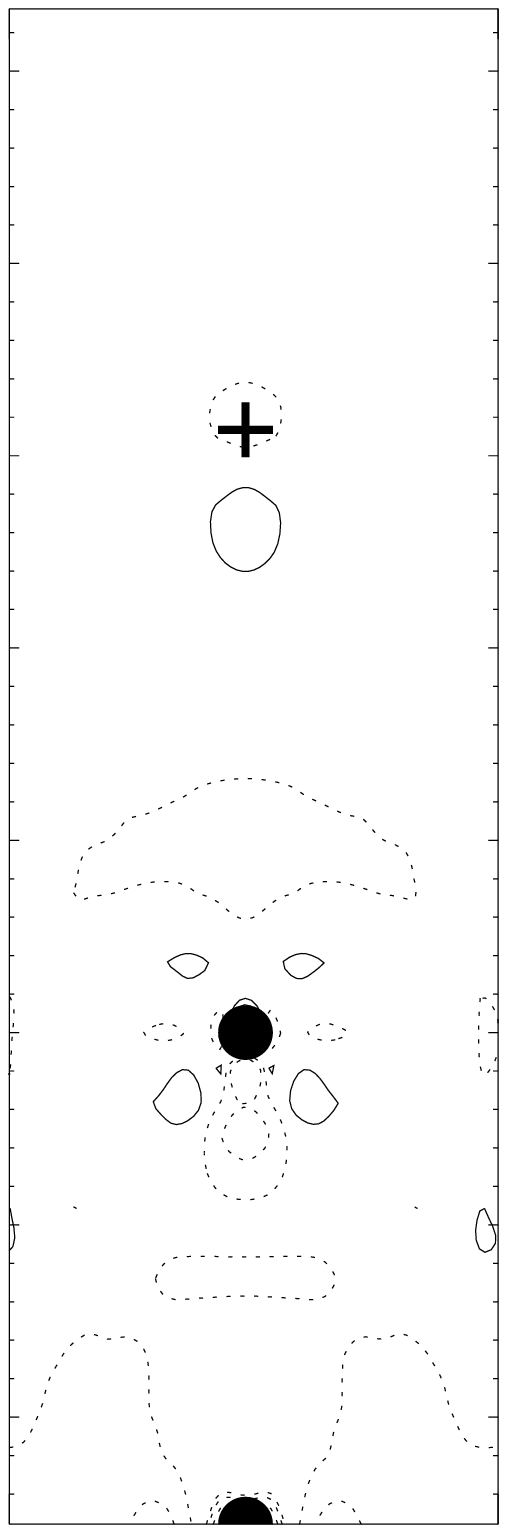,height=7cm}
\hspace{-2cm}
\psfig{figure=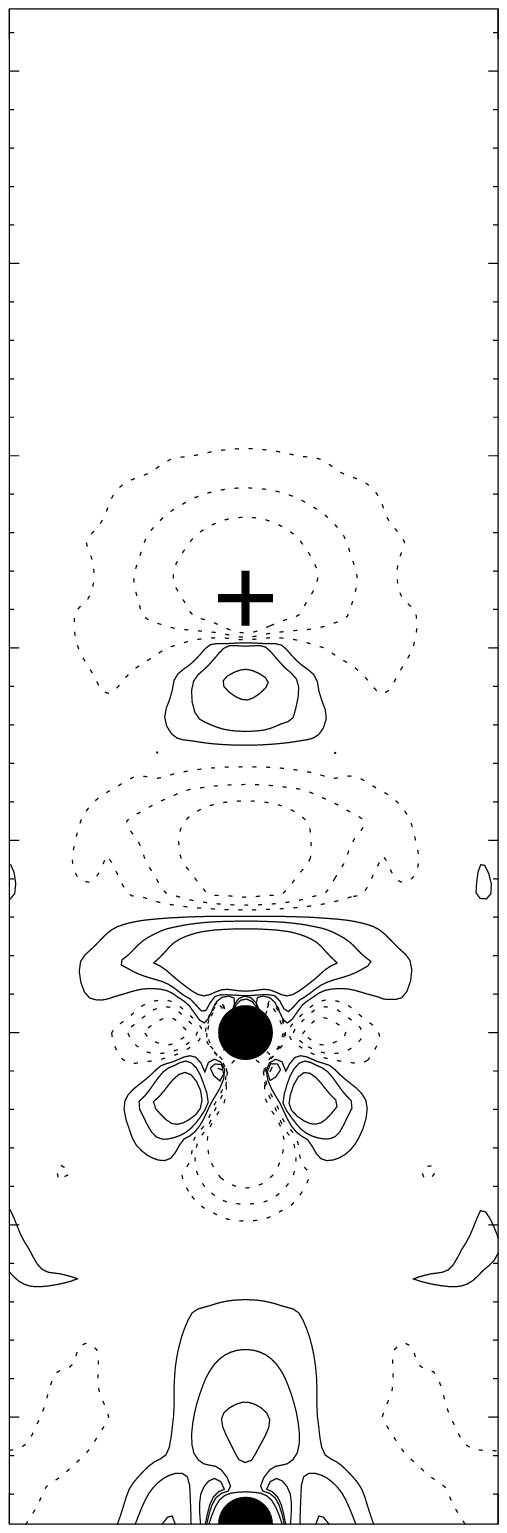,height=7cm}}}
\vspace{-1.2cm}
\centerline{\hbox{
\hspace{-1cm}
\psfig{figure=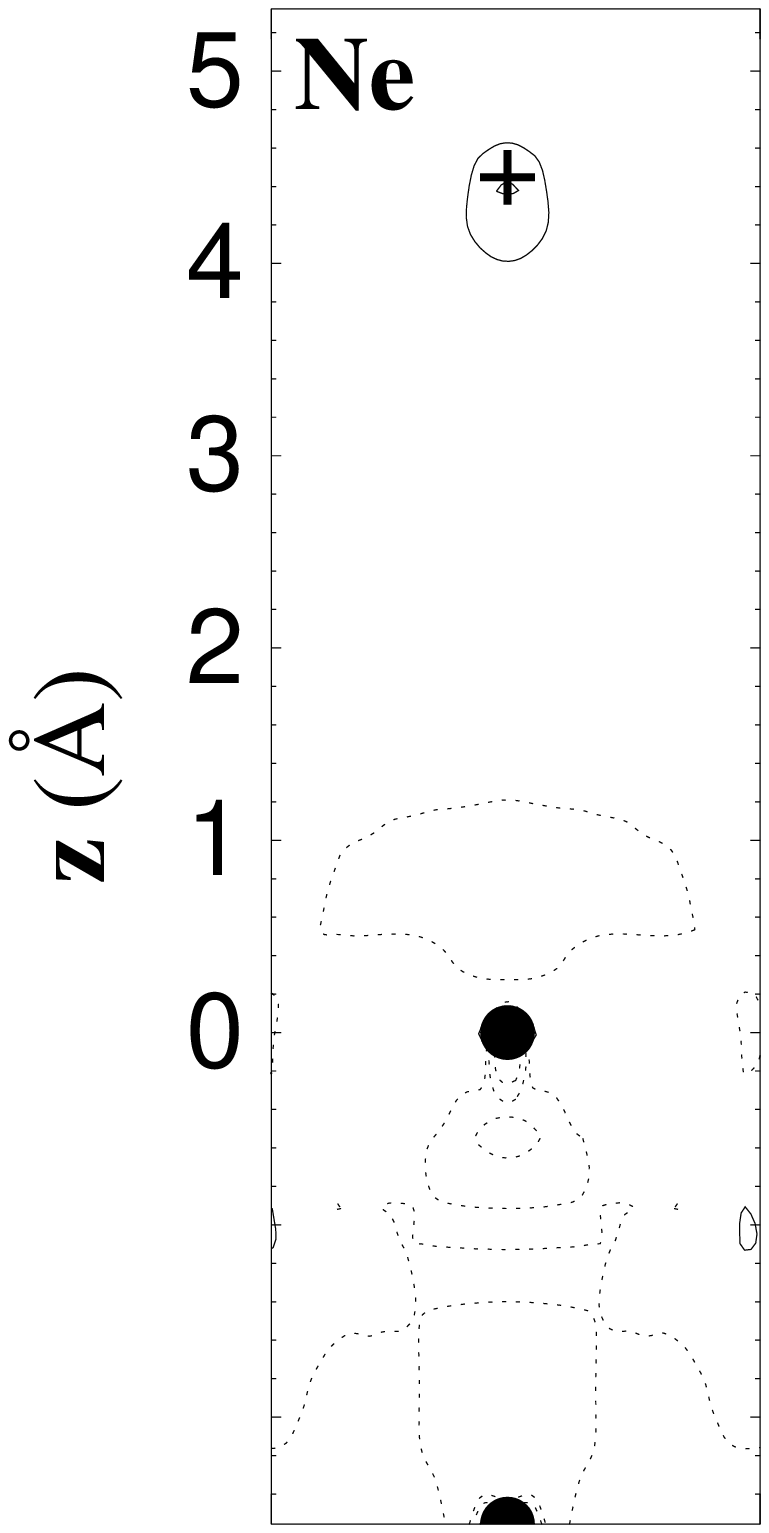,height=7cm}
\hspace{-2cm}
\psfig{figure=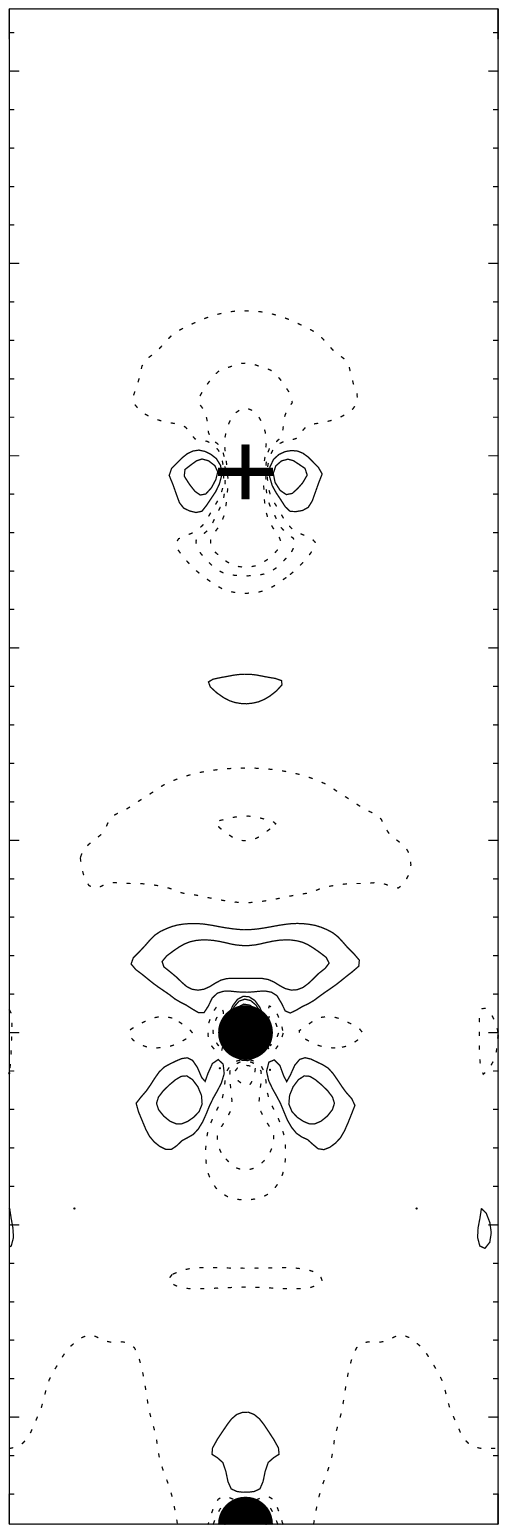,height=7cm}
\hspace{-2cm}
\psfig{figure=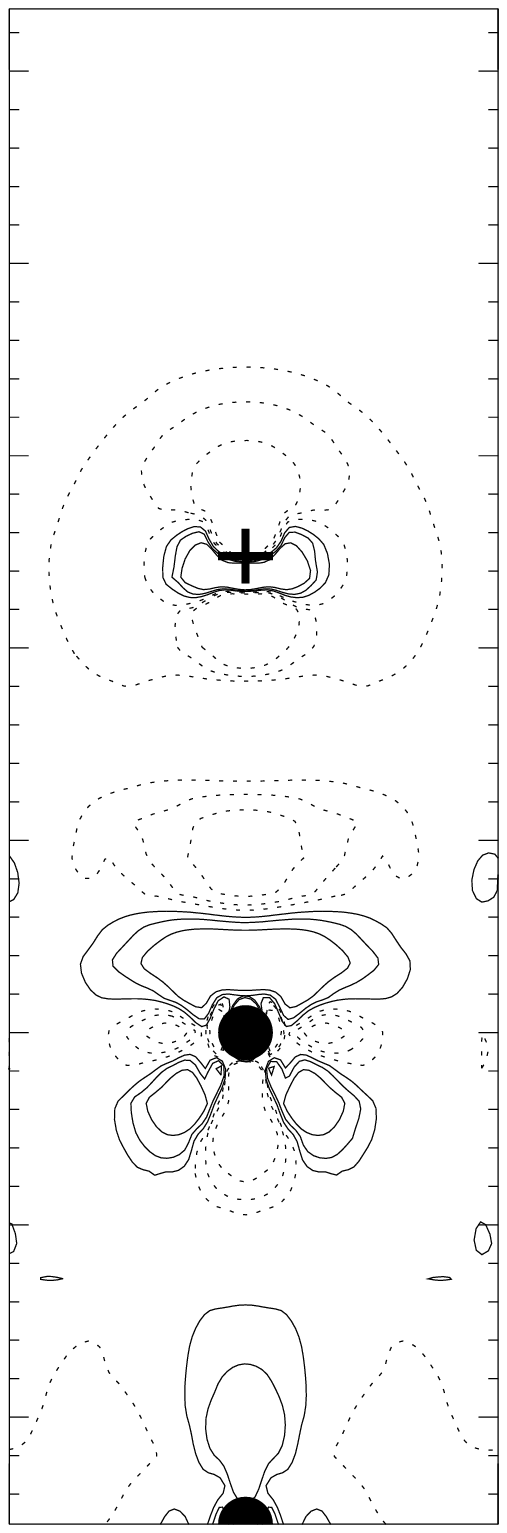,height=7cm}}}
\vspace{-.8cm}
\caption{Density difference plots (see text) for a He (top panels) and  Ne
(bottom panels)
atom over the on--top position at three different distances from the
topmost surface layer. The crosses indicate the positions of He (Ne) above
the surface. The distances are from left to
right: 4.2~\AA, 3.7~\AA~ and 2.2~\AA~ for He,
and 4.4~\AA, 3.7~\AA~ and 2.4~\AA~ for Ne.
Full lines note an increase and dashed lines a decrease in electron density;
the values are  $\pm 0.5, \pm 1.0, \pm 2.0$, and  $10^{-3}$~bohr$^{-3}$.}
\label{Densdiff}
\end{figure}

In order to analyze the differences of He and Ne
atom scattering we discuss the changes in the electron density of the
surface and of the rare-gas atoms induced by the interaction.
Figure~\ref{Densdiff} displays  the  difference between the self-consistent
electron  density of the interacting systems and the
superposition of the densities of the clean Rh\,(110) surface and a
He (Ne) atom. Three positions of the probe atoms are selected
which correspond to a slightly attractive interaction (left panel), to the
minimum of the interaction energy (middle panel), and to a position in the
repulsive part close to the turning point assuming a kinetic energy of 150 meV
(right panel). The figure shows clearly that both rare-gas atoms
change the substrate surface electron density noticeably
and that they are also significantly polarized themselves.
Thus, it is obvious that it is not the unperturbed surface electron density
which is probed by the scattering. Figure~\ref{Densdiff} shows
that for the surface the main changes occur in
the $d$ shell. The leading effects for both probe atoms are
a depletion of the $d_{3z^2 - r^2}$ states and an increase of
$d_{xz}$ and $d_{yz}$ occupation at the on-top position, and a depletion
of the $d_{xy}$ states and an increase of
$d_{xz}$ and $d_{yz}$ at the short--bridge site (The $x$ axis lies along the
short--bridge direction and the $y$ direction is parallel to the
long--bridge direction of the (110) surface).
The largest effect happens for the $d_{xz}$ electrons:
Compared to the unperturbed surface the $d_{xz}$ contribution
is increased at the turning point by  about 1 \%.
This increase is partially due to the fact that the Pauli repulsion
of the rare-gas atoms with the spilling out substrate $s$ electrons
is reduced by transferring $s$ electron density into the $d_{xz}$ band,
which for Rh has a particularly high density of filled and empty
states right at the Fermi level. We come back to the important role
of the $d_{xz}$ band below.

While the reaction of the substrate is similar for both probe
atoms, the polarization densities of the He and Ne atoms are clearly different
as is the nature of  their interaction with the surface.
For He the interaction
is mediated by the He $1s$ electrons and a polarization of the He atom away
from spherical symmetry which implies a hybridization
of $1s$ and $2p_{z}$ orbitals, clearly visible in Fig.~\ref{Densdiff}
(top right panel).
On the other hand, the interaction of Ne is dominated by
the  $2p$ electrons and the easier polarization of the Ne atom
which requires a $2p \rightarrow 3s$ virtual transition.
This contradicts the interpretation of Rieder {\it et al.} \cite{Rie94}
which was based on a strong involvement of the
Ne $2s$ orbitals.
As our calculations show (see also Fig.~\ref{Densdiff})
at the on--top position He exhibits a
reduction of the $1s$ density and an increase of the
$2p$ density (mainly $2p_{z}$); a result similar to that found
in studies of He physisorption at a jellium surface \cite{Lang}.
On the other hand, for Ne we
find a reduction of the $2p_{z}$ occupation and a slightly
stronger localization of the $p_{x}, p_{y}$ states.
At the short--bridge site the polarization of the He atom is similar to that
at the on--top geometry, although
weaker, but that of Ne is different, i.e., here the occupancy of all
three $2p$ states is reduced.

The results are understood as follows. The reflection
of He and Ne atoms happens rather close to the surface, at a distance
slightly less than 3 \AA, i.e., closer than the position of the physisorption
well.
Here the DFT-GGA approach describes the interaction
with sufficient accuracy.
The nature of the interaction is determined by electron polarizations
and hybridizations. Thus, it is not the {\em total} electron density
of the substrate surface which determines the interaction, but the
electronic wave functions which lie close to the Fermi level. The He $1s$
orbital
and the Ne $2p$ orbitals interact with these substrate states in a
qualitatively different manner.
For Rh\,(110) the states which are most important at the distance
of reflection have $d_{xz}$ character, because these
substrate states give rise to a very flat band (thus high density)
which crosses the Fermi level close to the Brillouin zone boundary. In other
words, these orbitals change phase when going from
one Rh atom to the next one along the short--bridge direction.
Thus, the Bloch state is bonding in character, although, due to the weak
overlap (reflected by the band's flatness) one may call it non bonding.

At the short--bridge position these $d_{xz}$ substrate states
are thus symmetric with respect to  mirror planes along $xz$ and $yz$, and
their electron density is low.
A He atom, with its $1s$ state, feels a Pauli repulsion with these states
and is efficiently reflected.
On the other hand, for Ne the $2p_x$ and $2p_{y}$ states are antisymmetric.
Thus, at the short--bridge position they will not interact with the $d_{xz}$
band. Only the Ne $2p{_z}$ orbital has the same symmetry, but due to its
narrow lobe and the low density of the substrate states at the
short--bridge position, it can still approach  the surface rather close
before the Pauli repulsion becomes important.

The opposite situation occurs for the on--top geometry were
the He $1s$ state is orthogonal to the substrate $d_{xz}$ wave functions
and can thus approach the surface quite close up to the point were the
repulsion with the energetically lower lying
$d_{3z^2-r^2}$ states dominates. Some of this repulsion is removed by
transferring $d_{3z^2-r^2}$ electrons into the $d_{xz}$ and
$d_{yz}$ bands.
For Ne the repulsion at the on-top site is much stronger: The  $2p_z$
electrons interact repulsively with the substrate $d_{{3z^2 - r^2}}$ electrons,
and the Ne $2p_{x}$  interact repulsively with the
$d_{xz}$ electrons. The Ne $2p_y$ orbital is found to be  affected only little.

These results imply that the interaction between rare gas atoms and a
surface is significantly more complicated that hitherto assumed: It is not the
total {\em electron density} of the surface which is probed, but the
interaction is determined by the substrate surface {\em wave functions} at the
Fermi level.
Although this is more complicated it is also more interesting, because these
states are important also for the chemical reactivity of the surface.
Our explanation of the interaction mechanism has interesting
consequences. For example, we expect similar ``anticorrugation'' effects
for the $d$ metals
which belong to the same  or a direct neighbor column of the periodic table,
but for systems with a different band structure we expect different effects.

We thank K.~H.~Rieder for stimulating discussions. The
work was partially supported by the Deutsche Forschungsgemeinschaft,
Sonderforschungsbereich 290.

\end{document}